\documentclass[]{mnras}
\usepackage{graphicx}	
\usepackage{amsmath}	
\usepackage{amssymb}	
\usepackage{bm}
 \def\Msun{M_\odot} \def\deg{^\circ} \def\/{\over}\def\kms{km s$^{-1}$}  \def\sin{{\rm sin}\ }     \def\be{\begin{equation}} \def\ee{\end{equation}} \def\({\left(} \def\){\right)} \def\[{\left[} \def\]{\right]} \def\sech{{\rm sech}}  \def\pr{p_r}\def\d{\partial}\def\cot{{\rm cot}} \def\Alf{Alfv{\'e}n }  
\def\Vu{V_{\rm unit}} \def\rhou{\rho_{\rm unit}} \def\Bu{B_{\rm unit}} \def\tu{t_{\rm unit}} \def\Va{V_{\rm A}} \def\cs{c_{\rm s}} 
\def\sech{{\rm sech}} \def\cosh{{\rm cosh}} 
\def\cs{c_{\rm S}}
\def\red{}
\title[Feedback between Sgr A and B]{Feedback between Sgr A and B :\\
AGN-Starburst Connection in the Galactic Centre} 

\author[Y. Sofue]{ Yoshiaki Sofue\thanks{E-mail: sofue@ioa.s.u-tokyo.ac.jp} \\ 
Institute of Astronomy, The University of Tokyo, Mitaka, Tokyo 181-0015, Japan }

\date{Accepted; Received YYY; in original form} 
\pubyear{2020}
 
\begin{document} 
\maketitle

\begin{abstract}  
Propagation of fast-mode magneto-hydrodynamic (MHD) compression waves is traced in the Galactic Centre. MHD waves produced by the active Galactic nucleus (Sgr A) focus on the molecular clouds such as Sgr B in the central molecular zone, which will trigger star formation, or possibly starburst. MHD waves newly excited by the starburst propagate backward, and focus on the nucleus (Sgr A), where implosive waves compress the nuclear gas to promote fueling the nucleus and may trigger nucleus activity. Echoing focusing of MHD waves between Sgr A (active galactic nucleus: AGN) and Sgr B (starburst) trigger each other at high efficiency by minimal energy requirement. It also solves the problem of angular momentum for AGN fueling, as the focusing waves do not require global gas flow. 
\end{abstract}  

\section{Introduction}  


AGN (active galactic nuclei)-starburst (SB) connection/link has been suggested based on correlation analyses between the luminosity of galactic nuclei and that of circum-nuclear warm dust heated by active star formation (SF)
 \cite{Smith+1998,Wild+2010,Fabian2012}.
Large amount of energy released at AGN produces various types of energetic outflows such as jets, bubbles and winds expanding into the galactic disc and halo
\cite{King+2015},
which interact with the circum-nuclear disc (CND) and torus as well as gas clouds, and enhance star formation (SF) and starburst.
This is categorized as the AGN-to-SB feedback, which is the first subject of this paper.

On the other hand, the surrounding gas disc plays a role in fueling the gas to the nucleus by overcoming the problem of the refusing force due to the conservation of angular-momentum by
(i) bar-dynamical
\cite{Shlosman+1989},
(ii) magnetic braking
\cite{Krolik+1990},
and
(iii) radiation drag
\cite{Umemura+1997,Thompson+2005} 
accretion mechanisms.
These may be categorized as disk-to-AGN fueling.
However, feedback of (iv) star formation and/or starburst itself to the AGN has not been thoroughly investigated, which is the second subject of this paper.


In our Galactic Centre, various expanding and out-flowing phenomena have been observed, indicating that the Milky Way has experienced AGN phases in the past with various energies and time scales.
Expanding phenomena are evidenced, for example, by multiple thermal shells of radii $\sim 10$ pc around Sgr A with required energy of $\sim 10^{51}$ ergs in the last $\sim 10^5$ y 
\cite{Sofue2003},
200-pc expanding molecular cylinder of $\sim 10^{54}$ ergs
\cite{Kaifu+1972,Scoville1972,Sofue2017},
GC radio lobe of $\sim 200$ pc at $\sim 10^{54}$ ergs 
\cite{Sofue+1984,Heywood+2019},
and giant shells/bubbles in the halo in radio, X-rays and $\gamma$-rays with radii from several to $\sim 10$ kpc with $\sim 10^{55}$ ergs in the last $10^6$ to $10^7$ y
\cite{Sofue1980,Sofue2000,Sofue+2016,Su+2010,Crocker2012,Kataoka+2018}.
Numerous non-thermal filaments in radio continuum emission 
\cite{YZ+2004,LaRosa+2005,Lang+1999} may also indicate continuous magneto-hydrodynamic (MHD) waves excited by the activity in Sgr A 
\cite{Sofue2020a}.

The Galactic nucleus is surrounded by the central molecular zone (CMZ) embedding active star forming regions such as Sgr B and C
\cite{Morris+1996,Oka+1998,Oka+2012,Tsuboi+2015}. 
High excess of the number of supernova remnants (SNR) in the GC direction indicates a high rate of SF in the CMZ
\cite{Gray1994}.  
Star formation in the CMZ has been discussed often in relation to the non-linear response of the rotating disc gas to the barred potential 
\cite{Krumholz+2015} and 
to cloud-cloud collisions 
\cite{Hasegawa+1994,Tsuboi+2015}.
Feedback of the nuclear activity 
\cite{Zu2015,Zu+2013,Zu+2017,Hsieh+2016}
would also trigger the SF in CMZ. 

The SF activity would in turn disturb the surrounding medium by supernova explosions and stellar winds
\cite{Martinpintado+1999}.
Radio continuum blobs and filaments as mixture of thermal and non-thermal emissions, composing a radio-bright zone (RBZ), suggests high-energy feedback from the SF regions to the CMZ and surrounding medium
\cite{Zhao+2016,Yusef+2019}.   
Thus produced disturbances will propagate through the RBZ, and further reach and affect the nuclear region around Sgr A.
However, such feedback from the SF regions to Sgr A seems to be not thoroughly investigated.

In this paper, we investigate the propagation of MHD waves in the Galactic Center by solving the Eikonal equations for low amplitude fast-mode MHD waves in magnetized medium by the method described in section \ref{secmethod}. 
In section \ref{secsgrA} MHD disturbances induced by the activity in Sgr A are shown to converge on the CMZ and molecular clouds therein to compress and trigger SF, which may hint to insight into an efficient, and hence minimal energetic feedback in the AGN-SB connection.
In section \ref{secsgrB} we trace the waves emitted from the SF region, and show that they converge onto the nucleus at high efficiency, which will trigger AGN activity in Sgr A. In section \ref{secdiscussion} we discuss the implication of the result.

Throughout the paper, the term "Sgr A" will be used to express the complex around the Galactic nucleus including Sgr A$^*$ (AGN of the Milky Way) and associated molecular and radio sources.
Similarly 'Sgr B' expresses the SF region and molecular complex associated with the radio sources Sgr B1 and B2, which is embedded in the CMZ and is supposed to be a starburst site.
 
\section{Method} 
\label{secmethod}
\subsection{Basic equations}

Disturbances excited by an explosive event in the interstellar medium  propagate as a spherical shock wave in the initial phase. In the fully expanded phase, they propagate as sound, \Alf, and fast-mode MHD waves. Among the modes, sound wave is much slower than the other two modes in the usual ISM condition. The \Alf wave transports energy along the field lines, while it does not compress the field, so that it is not effective in compressing the gas to trigger star formation. The fast-mode MHD wave (hereafter, MHD wave) propagates across the magnetic field lines at \Alf velocity, and compresses the local field as well as the gas, which would act to trigger star formation
\cite{Sofue2020a}.

The basic equations of motion, or the Eikonal equations, to trace the fast-mode MHD waves were obtained in order to study the Morton waves in the solar corona 
\red{under the condition that the \Alf velocity is sufficiently higher than the sound velocity, or in the so-called low $\beta$ condition.}
\cite{Uchida1970,Uchida1974}. 
The method has been applied to MHD wave propagation in the Galactic center, supernova remnants, and star forming regions 
\cite{Sofue1978,Sofue1980,Sofue2020a,Sofue2020b}. 
Given the distribution of \Alf velocity and initial directions of the wave vectors, the equations can be numerically integrated to trace the ray path as a function of time. 
The equations are shown in the Appendix as reproduced from the above papers. 

Besides galactic disc at rest, we also examined a case that the disc is rotating at a constant rotation velocity.
Thereby, the azimuth angle moment of each wave packet,  $p_\phi$ in the Eikonal equations, was modified by adding the angular velocity caused by the rotation around the $z$ axis. 

Numerical integration of the differential equations was obtained by applying the first order Runge-Kutta-Gill (RKG) method with sufficiently small time steps.
The validity was confirmed by checking some results by applying the second order RKG method as well as by changing the time steps.
Because of the simple functional forms of the adopted \Alf velocity distributions having no singularity, the 1st order method was sufficiently accurate and faster than the 2nd order method.
While the computations were performed in the spherical coordinates, the results will be presented in the Cartesian coordinates $(x,y,z)$, where $x$ and $y$ represent the distance in the galactic plane, $z$ is the axis perpendicular to the disc with $(0,0,0)$ denoting the nucleus, and $\varpi=(x^2+y^2)^{1/2}$ is the distance from $z$ axis.

\subsection{Gas distribution}

Extensive observations in molecular line, radio and X-ray observations have revealed that the gas density distribution is expressed by a disk-like CMZ surrounded by a molecular ring 
\cite{Morris+1996,Oka+1998,Sofue1995,Sofue2017},
warm gas disk
\cite{OkaTake+2019},
and extended hot gas 
\cite{Koyama2018}.
We represent the gas distribution by superposition of several components:
\be
\rho=\Sigma_i \rho_i,
\ee
where individual components are given as follows:
The main disc is represented by
\begin{equation} 
\rho_{\rm disk}= \rho_{\rm disk,0}
\sech \left(\frac{z}{h}\right) 
e^{-(\varpi/\varpi_{\rm disk})^2}.
\end{equation} 
We adopt $\rho_{\rm 0,disk}=1.0$ and $\varpi_{\rm disk}\sim 10$ in the scale units as listed in table \ref{tab_unit} (described later).
 
A molecular ring representing the main body of CMZ is represented by
\be
\rho_{\rm ring}= \rho_{\rm 0,ring}
e^{-((\varpi-\varpi_{\rm ring})^2+z^2)/w_{\rm ring}^2},
\ee 
where $\varpi_{\rm ring}=5$ and half ring width of $w_{\rm ring}\sim0.5-1$. 
Gas clouds are assumed to have Gaussian density distribution as
\be
\rho_{\rm cloud,i}=\rho_{0, i} 
e^{-(s_i/a_i)^2}, 
\ee
where $s_i^2=(x-x_i)^2+(y-y_i)^2+(z-z_i)^2$, $\rho_{\rm cloud, i, 0}$ and $a_i$ arei centre density and scale radius of the $i-$th cloud or gaseous core centered on $(x_i,y_i,z_i)$.
The whole system is assumed to be embedded in a halo of 
\be
\rho_{\rm halo} =0.01.
\ee
As a nominal set, we take $\rho_{\rm coud, i}=100$, $a_i=1$, $\rho_{\rm disk, 0}=1$, $h=1$.
 
\subsection{Magnetic fields}

Non-thermal radio emission in the GC is more extended than the molecular gas disc and clouds, indicating that the magnetic pressure distribution is smoother than the gas distribution and the field strength is on the order of $0.1-1$ mG.
There may be two major components.
One is the large-scale vertical/poloidal field penetrating the galactic disc with roughly constant strength at $\sim 0.1-1$ mG
\cite{YZ+1987,Tsuboi+1986,Sofue+1987}, 
and the other is a ring field of radius $\sim 100-200$ pc, whose strength is $\sim 0.01-0.1$ mG
\cite{Nishiyama+2010}.

\red{The strong magnetic field in the GC of 0.1 to 1 mG may be explained by a primordial-origin model, in which the primordial magnetic field was gathered in the GC during the the proto-Galactic accretion to form a strong vertical field  \cite{Sofue+2010}.
The field strength is amplified to a value at which the magnetic energy density balances the kinetic energy density of the disc gas in galactic rotation at $\sim 200$ \kms in the deep gravitational potential of the GC. }

In the galactic disc of solar vicinity, Zeeman effect observations in local molecular clouds indicate that the magnetic strength is about constant at several $\mu$G through molecular clouds with density less than $\sim 10^4$ H cm$^{-2}$ except for high-density cores \cite{Crutcher+2010}, and the \Alf velocity decreases with the gas density \cite{Sofue2020b}.
We here also assume such a general property of magnetized clouds in the GC.

In our simulation, we first examine simple cases assuming a constant magnetic field $B=B_{\rm halo}=1$ in order to show  typical behaviors of wave propagation.
Then, we adopt a more realistic magnetic fields, where the fields are loosely coupled with the gas distribution in such that the magnetic pressure varies with with scale radii being twice those for the gas distribution, namely
\be
B_{\rm disk}^2= B_{\rm disk,0}^2 
\sech \left(\frac{z}{2h}\right) 
e^{-(\varpi/(2\varpi_{\rm disk}))^2}
\ee
in the disc, and 
\begin{equation}
    B_i^2=B_{i,0}^2 
    e^{-(s_i/(2a_i))^2}
\end{equation}  
in the ring and clouds, where $s_i$ is the distance from the center of $i$-th cloud or ring's azimuthal axis.

We further examine a case that the disc is penetrated by a vertical cylinder of magnetic flux.
In our recent paper 
\cite{Sofue2020b} we modeled the vertical radio continuum threads
\cite{Heywood+2019} as remnants of nuclear activity.
Thereby, we assumed a large-scale vertical magnetic cylinder penetrating the disc following the primordial origin model of Galactic magnetic field
\cite{Sofue+2010}.
It was shown that the MHD waves are confined inside the magnetic cylinder, forming vertically stretched filaments, which well reproduced the observed radio threads. 
The magentic cylinder is represented by
\be
B_{\rm cyl}=5 \ e^{-(\varpi-\varpi_{\rm cyl})^2/w_{\rm cyl}^2},
\ee
where $\varpi_{\rm cyl}=3$ and $w_{\rm cyl}=1$.

\subsection{\Alf velocity}

The \Alf velocity is given by
\be
\Va=\sqrt{\Sigma_i B_i^2/{4\pi \rho}},
\ee 
where $i$ represents each of components of the interstellar medium in the GC.
\red{It is estimated to be 
$\Va \sim 70-700$ \kms in the GC disc with $B\sim 0.1-1$ mG and $\rho\sim 10$ H cm$^{-3}$.
On the other hand, the sound velocity in the disc is $\cs \sim 1$ \kms for HI gas of temperature of $T\sim 100$ K and $\sim 0.3$ \kms for molecular gas of $\sim 20$ K. 
In the molecular ring with $B\sim 0.1$ mG, $\rho\sim 100$ H cm$^{-3}$ and $T\sim 20$ K, we have $\Va \sim 20$ \kms and $\cs \sim 0.3$.
In dense molecular clouds with $B\sim 0.1$ mG, $\rho\sim 10^4$ H cm$^{-3}$ and 10 K, we have $\Va \sim 2$ \kms and $\cs \sim 0.3$ \kms. 
The \Alf velocity in the halo increases rapidly above/below the gaseous disc, whose scale height is much smaller than that of magnetic field, so that $\Va \gg \sim 100$ \kms, the sound velocity of the X-ray halo at $\sim 10^6$ K.
Thus, we may safely assume that the \Alf velocity is sufficiently faster than the sound velocity in the circumstances under consideration, which was the condition for the Eikonal method used in this paper.
}

Besides such strong global fields, we may also consider other possible models as follows. 
If the magnetic field and interstellar gas are in a local energy-density (pressure) equipartition, $B^2 \propto \rho \sigma_v^2$, the \Alf velocity is nearly equal to the turbulent velocity $\sigma_v$ of the gas, which is usually almost constant at $\sim 5- 10$ \kms.
In such a case of constant \Alf velocity, the MHD waves propagate rather straightly without suffering from deflection.
On the other hand, if the magnetic field is frozen into the gas, the magnetic flux is proportional to $\rho^{2/3}$, or $\Va \propto \rho^{1/6}$.
This means that the \Alf velocity increases toward the cloud center, and the cloud works to diverge the waves rather than to converge. 
Such a case is indeed observed in high-density molecular cores with density $\sim 10^5$ H cm$^{-3}$ \cite{Crutcher+2010}. 
\red{These cases may apply to individual clouds, not largely changing the global wave propagation in the GC, and will be considered as a cause for local fluctuations as discussed in section \ref{clumpiness}.}
 
\begin{table} 
\begin{center}
\caption{Units used in the calculation}  
\label{tab_unit}  
\begin{tabular}{ll}  
\hline   
Density, $\rhou=\rho_{\rm disk, 0}$ & 100 H cm$^{-3}$ \\
Magnetic field, $\Bu$ & 1 mG \\
Velocity, $\Vu =\Bu/\sqrt{4\pi \rhou}$ & 220 \kms \\
Length $A$& 40 pc\\
Time, $\tu=A/\Vu$ & 0.18 My \\
\hline   
\end{tabular} 
\end{center}
\end{table} 

\subsection{Units and normalization}

The real quantities are obtained by multiplying the units to the non-dimensional quantities in the equations, where the length by $A$, time by $\tu$, and velocity by $\Vu =\Bu/\sqrt{4\pi \rhou}=219$ \kms as listed in table \ref{tab_unit}, following our recent analysis of the GC threads 
\cite{Sofue2020b}. 
Typical \Alf velocity is $\Va \sim 20$ \kms in molecular clouds and $\sim 200$ \kms in the galactic disc. 

\subsection{Dissipation rate}

The dissipation rate $\gamma$ of a small-amplitude MHD wave defined through amplitude $\propto \exp(-\gamma L)$ 
\cite{Landau+1960} is expressed as
\be
\gamma=\frac{\omega^2}{2V^3} \(\frac{\nu}{\rho} + \frac{c^2}{4\pi\sigma_e}\),
\ee
where $\omega$ is the frequency,  $\nu\sim 10^{-4}$ g cm s$^{-1}$ is the viscosity of hydrogen gas, $\sigma_e$ the electric conductivity, $c$ the light velocity, and $L$ is the distance along the ray path. The first term is due to dissipation by viscous energy loss, and the second term due to Ohmic loss, which is small enough compared to the first. 
Then, the dissipation length $L$ is estimated by
\begin{eqnarray} 
\(\frac{L}{{\rm kpc}}\)=\gamma^{-1} \sim 0.6 
\(\frac{B}{{\rm \mu G}}\) 
\(\frac{\rho}{ {\rm H \ cm^{-3}} } \)^{1/2}\nonumber \\
\times \(\frac{\nu}{{\rm 10^{-4} g \ cm \ s^{-1}}}\)^{-1} 
\(\frac{\lambda}{{\rm pc}}\), 
\label{dissipation}
\end{eqnarray} 
which amounts to several kpc in the region under consideration, so that the dissipation is negligible in the GC region. 

\subsection{Assumptions and limitation of the method} 

The method here assumes that the amplitude of the MHD waves is small enough and the Eikonal equations are derived using the linear wave approximation.
Hence, non-linear compression of the interstellar medium and the real density and magnetic field in the waves cannot be calculated in this paper.
Therefore, we here discuss only the possibility of mechanisms to compress the gas and to enhance or trigger the star formation and nuclear activity from the relative amplification of wave amplitude due to the geometrical effect of spherical implosion of the wave front onto the focal point.

 Although the direction of the propagation does not depend on the magnetic field direction, the compression by the fast-mode MHD wave occurs in the direction perpendicular to the magnetic lines of force, with the amplitude proportional to \sin $i$, where $i$ is the angle between the direction of the propagation and magnetic field.
This implies that the gas compression is not effective when the wave direction is parallel to the field ($i \sim 0$), whereas it attains maximum at perpendicular propagation ($\sim 90\deg$).

In the GC, the global field direction is observed to be perpendicular to the galactic plane in the inter-cloud and out-of plane regions \cite{Heywood+2019}, while it is nearly parallel to the molecular ring occupying most of the CMZ \cite{Nishiyama+2010}.
Therefore, the waves expanding from Sgr A toward the molecular ring and those from the CMZ toward the nucleus propagate almost perpendicularly to the global magnetic fields at maximum or high compression efficiency.

Magnetic structures in the molecular clouds and the nuclear regions are not well observed, so that we here assume that they are random.
In such a case, the field directions can be assumed to be statistically oblique at the most expected angle of $i\sim 60\deg$ (angle dividing a sphere into two equal areas about the polar axis), so that the compression efficiency is $\sim {\rm sin} \ 60\deg = 0.87$ as a mean. 
Thus, we may here consider that the fast-mode MHD waves propagate at high compression efficiency in almost everywhere in the GC for the first approximation.

\section{Sgr A to B: AGN to CMZ} 
\label{secsgrA}

	\begin{figure}
\begin{center}    
\includegraphics[width=7cm]{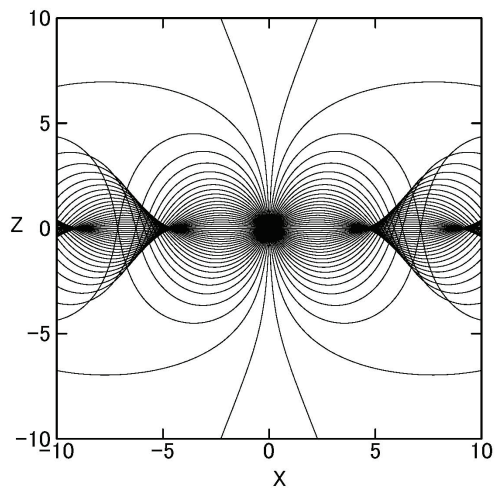}      
\end{center}
\caption{Ray paths of MHD waves from the galactic centre through a sech disc in the $(x,z)$ plane.} 
\label{rays_disk} 
	\end{figure} 
 
	\begin{figure*}
\begin{center}    
\includegraphics[width=16cm]{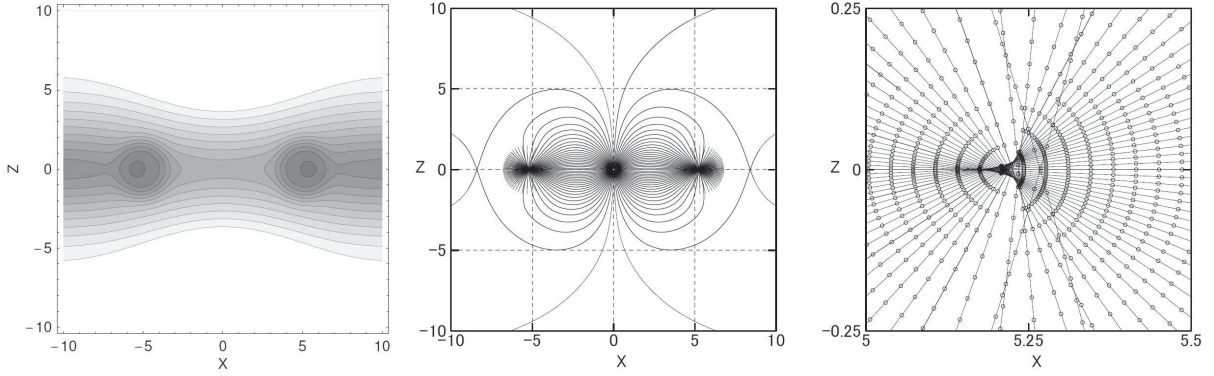}       
\end{center}
\caption{(Left) Density distribution in the $(x,z)$ plane. Dark: log $V/V_{\rm unit}=-1$, white: 1. (Middle) MHD wave ray paths. (Right) Enlargement near the focus (ring).} 
\label{Valf}  
\label{rays_ring} 
	\end{figure*} 
        
	\begin{figure} 
\begin{center}    
\includegraphics[width=7cm]{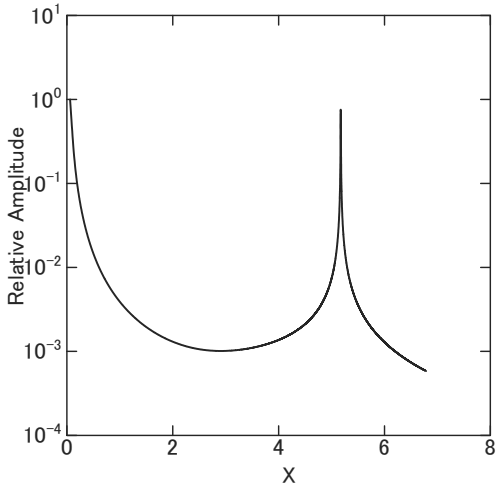}  
\end{center}
\caption{MHD wave amplitude.} 
\label{flux} 
	\end{figure}

	\begin{figure*} 
\begin{center}   
{\bf [Sgr A $\Rightarrow$ Disc and ring]} \\
\includegraphics[width=13cm]{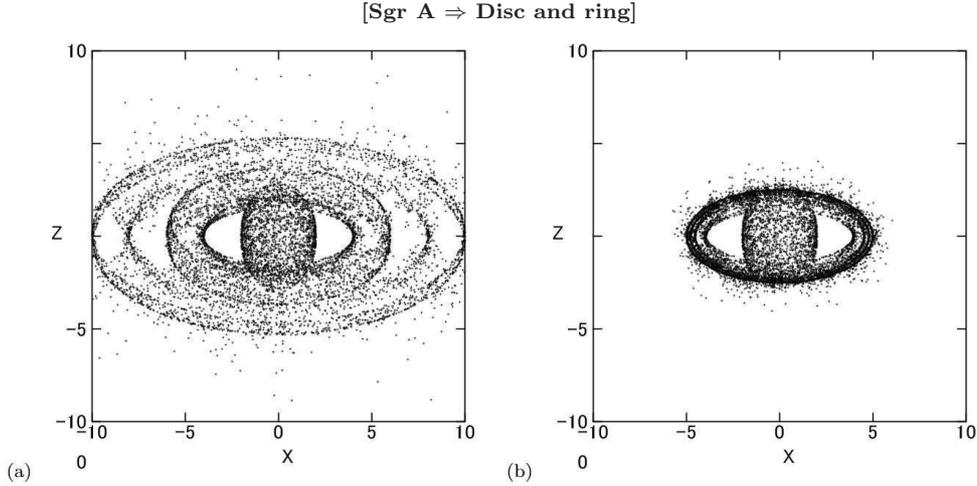}  
\end{center}
\caption{(a) MHD waves from the nucleus propagating through the sech h gas disc with constant magnetic field. (b) Same converging on a gaseous ring with $r=5$ and width $r_{\rm ring}=0.5$ with parameters as
$\rho_{\rm disk}=1/\cosh(z/1.)$;
$\rho_{\rm ring}=100 \exp(-((r-r_{\rm ring})^2+z^2)/r_{\rm ring}^2)$; 
$rho=\rho_{\rm disk}+\rho_{\rm ring}$. 
 Each dot represents MHD wave packet whose propagation is traced by solving the Eikonal equations. Initially, about a thousand packets are put at random on a small sphere at the center with outward radial vectors. 
The ensemble of packets are plotted at a constant time interval, here every 2 units of time at $t=2$, 4, 6, ... .
The front expands spherically in the initial phase, elongated in the direction perpendicular to the disc, and reflected/refracted by the disc to focus on a focal ring (a).
If there is a molecular ring of radius $\varpi=5$, the waves focus more efficiently on the ring (b).  
}
\label{p-gc-ring} 
	\end{figure*} 

\begin{figure*} 
\begin{center}    
{\bf [Sgr A $\Rightarrow$ Sgr B, etc.]}\\
\includegraphics[width=12cm]{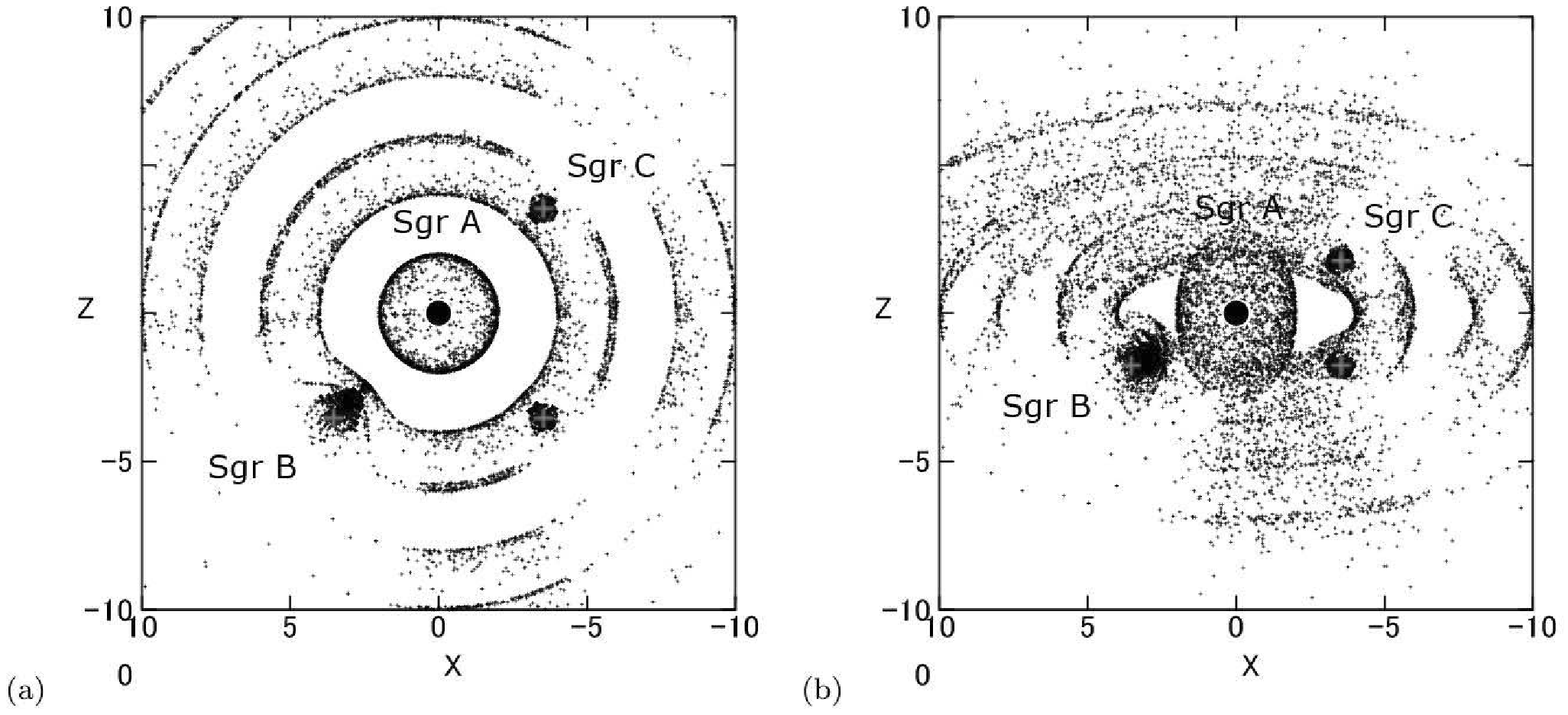}
\end{center}
\caption{(a) MHD waves from the nucleus converging onto three clouds projected on the xy plane.
(b) Same, but projected on a plane perpendicular to the line of sight at 30$\deg$ from the plane.  } 
\label{p-AtoB} 
	 
\begin{center}
{\bf [Sgr A $\Rightarrow$ Sgr B, etc. in rotation]}\\
\includegraphics[width=12cm]{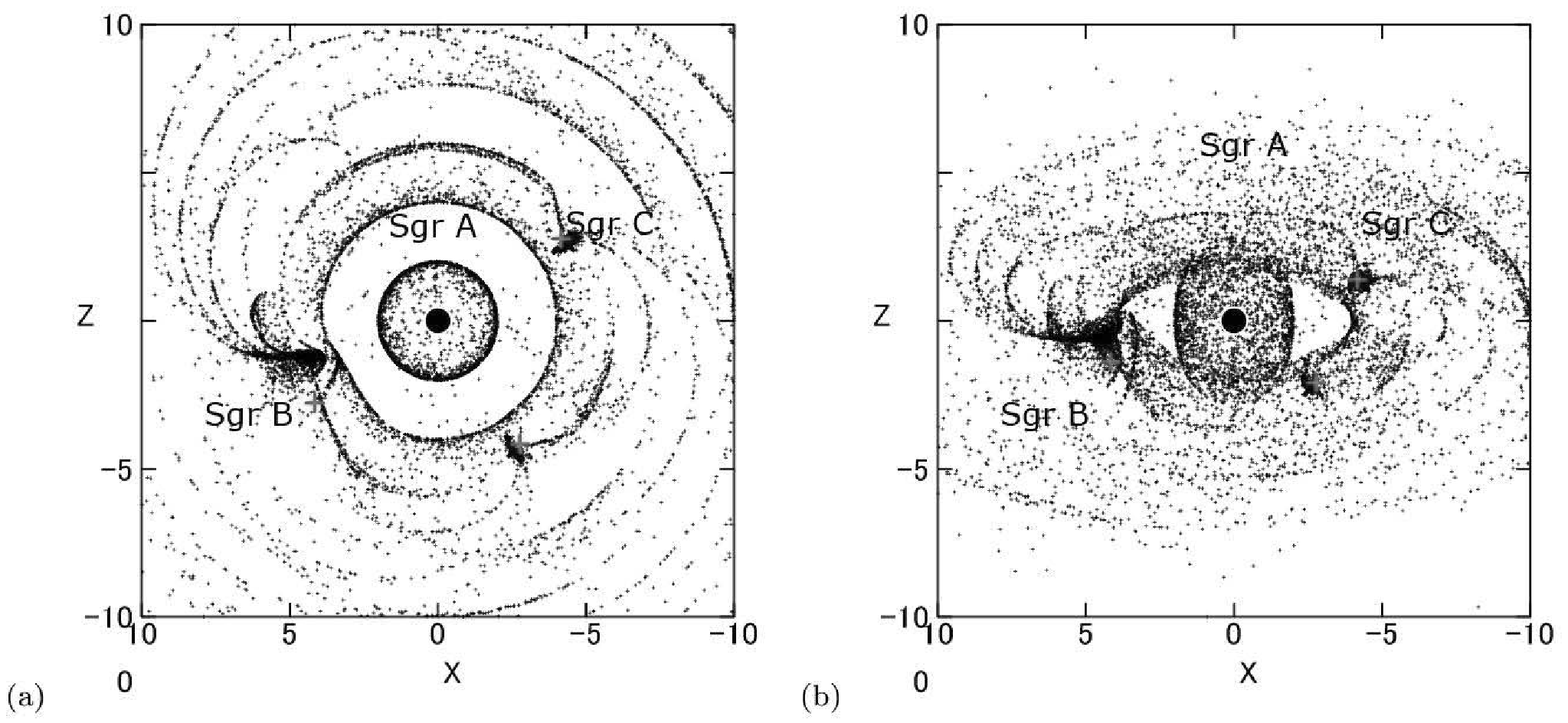}
\end{center}
\caption{Same as figure \ref{p-AtoB}, but the disc is rotating as figure \ref{Vrot}. } 
\label{p-AtoB-rot} 
	\end{figure*}

		\begin{figure}
\begin{center}    
\includegraphics[width=8cm]{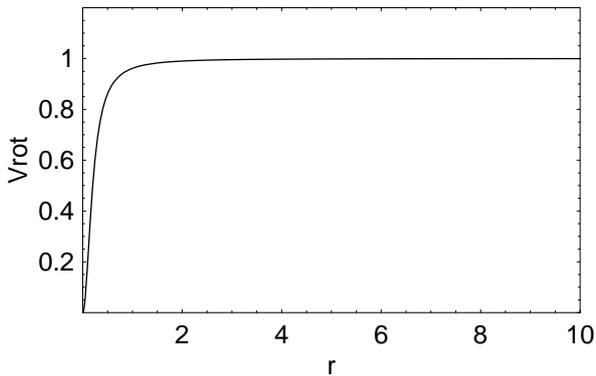}  
\end{center}
\caption{Rotation velocity of the disc and clouds.} 
\label{Vrot} 
	\end{figure}

	\begin{figure*}
\begin{center} 
{\bf [Sgr A $\Rightarrow$ Sgr B, etc. in rotation through vertical magnetic cylinder]}\\
\includegraphics[width=16cm]{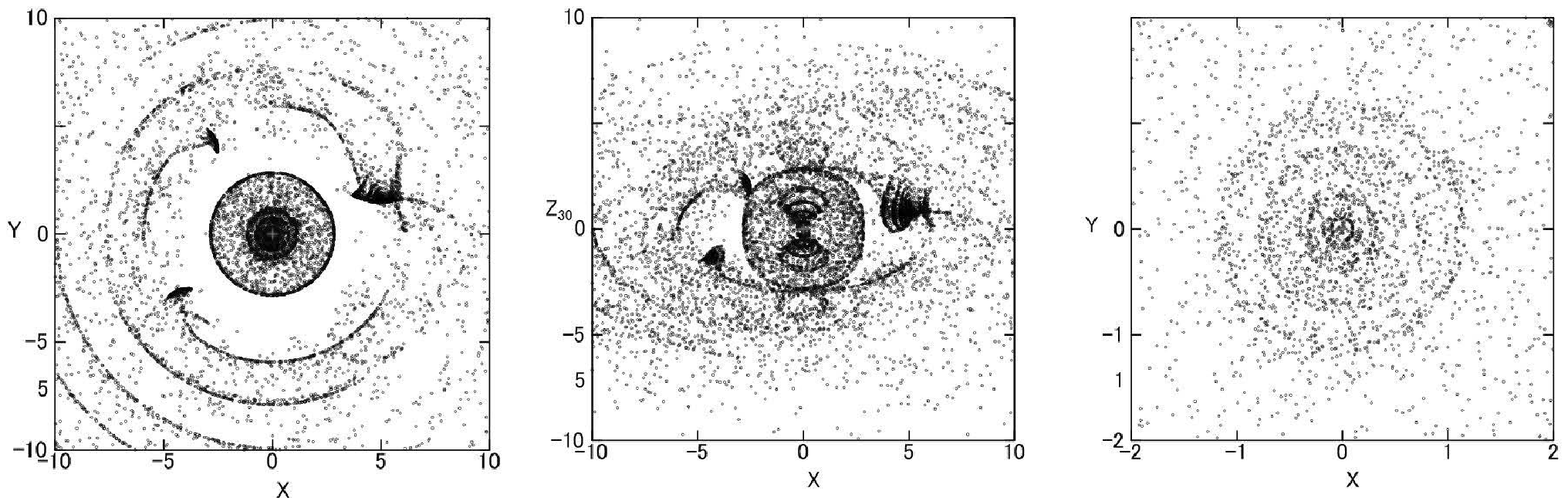}   
\end{center}
\caption{Same as figure \ref{p-AtoB-rot}, but a vertical magnetic cylinder of radius $\varpi_{\rm cyl}=3$ is present. MHD waves emitted at Sgr A are reflected and trapped inside the magnetic cylinder, while a significant fraction penetrates through it and converges onto the molecular clouds such as Sgr B. The right panel enlarges the central region to show that the reflected waves are converging back to Sgr A, indicating a self-feedback. Note a vacant area in the magnetic cylinder, from where the waves are rejected.}
\label{cylMag-AB} 
	\end{figure*}
   
\subsection{Propagation through the disc} 

We examine the effect of disturbances produced by explosive activity in the nucleus (Sgr A), which propagate and converge onto the CMZ and molecular clouds.
We first present a simple case of propagation through a plane-parallel layer with sech-type density profile, where the gas density has the form of
\be 
\rho_{\rm disk}=\sech (z/h),
\ee 
with $h=1$.
Figure \ref{rays_disk} shows the calculated result for the ray paths in the $(x,z)$ plane.

The MHD wave front expand spherically in the initial phase, and are reflected and refracted by the disc due to the rapid increase of \Alf velocity with the height.
The waves are then converged onto a circle in the plane (ring) with radius $\varpi \sim 4.4 h$. 
After focusing, the waves further propagate outward, and focus again at $r\sim 9h$, making the second focal ring. 
By such repetitive reflection and focusing, the waves are confined within the disc at high efficiency, so that the released energy is transformed outward, repeating circular and periodic convergence on the focal rings every $\sim 4.4h$ in radial interval.

\subsection{Convergence onto gaseous ring}

We next examine a case with a molecular gas ring of radius $\varpi_{\rm ring}=5$ around the nucleus embedded in the sech disk. 
We assume that the magnetic field is constant at $B=1$, and gas distribution is given by
\be 
\rho=\rho_{\rm disk}+\rho_{\rm ring} + \rho_{\rm halo},
\ee 
where
\be 
\rho_{\rm ring}=10 \ e^{-((\varpi-\varpi_{\rm ring})^2 +z^2)/w_{\rm ring}^2)},
\ee 
with $\varpi_{\rm ring}=5$ and $w_{\rm ring}=1$.

Figure \ref{Valf} shows the distribution of the \Alf velocity in the $(x,z)$ plane, where the ring has minimum \Alf velocity. 
The ray paths of waves propagating in such \Alf distribution are shown in the middle panel of figure \ref{rays_ring}, which indicates that the waves are strongly focused on the ring. 
The right panel enlarges the focal region, where the rays sharply focus on a circle of radius $\varpi=5.2$ slightly outside the gas ring's center at $x=5$. 

Figure \ref{flux} shows relative density of the wave packets, representing the wave flux, as calculated by $f=(z_0/z)^2$ for the rays near the galactic plane, where $z_0$ is the initial radius of the MHD wave front. 
This figure qualitatively represents the variation of wave flux as a function of the distance from the nucleus. 
In the figure, the wave flux is amplified at the focal ring by a factor of $\sim 10^3$, reaching almost the same flux as that in the initial sphere at the nucleus. 
However, in principle, because the rays of the wave packets cross eath other at the focus, where the area of the wave front becomes infinitesimally small ($z\simeq 0$), the amplification factor will reache infinity by the geometrical effect.

 In order to visualize the behavior of the wave front during the propagation, we adopt a dot plotting method. 
At the initial epoch, $t=0$, a number of wave packets, here about a thousand, are distributed at the origin with random radial vectors. 
Propagation of each of the wave packets is traced by the Eikonal equations, and the packets are displayed by dots projected on the sky at a given time interval. 

Figure \ref{p-gc-ring} shows thus calculated wave packets projected on a tilted plane by 30$\deg$ from the $(x,z)$ plane.
Each group of dots represents the wave front at a certain epoch. 
Here, the front is displayed every 2 time units, or at $t=2$, 4, 6, ..., 20.
The left panel shows a case for the sech disc (same as figure \ref{rays_disk}), where the wave front expands, reflected to converge on a focal ring at $r\sim 4$, again expands and focus on the outer focal ring at $r\sim 9$.
The right panel shows a front expanding into the disc superposed with a dense gaseous ring of radius $r=5$ and width $w=1$.
The waves behave in the same way as in the left panel inside the ring, but more strongly converge onto the gas ring, and are trapped there for a while.  

\subsection{Convergence onto clouds}

It is more likely that the CMZ is clumpy composed of molecular clouds. 
We examine a case when three molecular clouds with different sizes and densities are present at the same distance as the above ring embedded in the central disk.
We assume that the magnetic field is constant at $B=1$, and gas distribution is given by
\be 
\rho=\rho_{\rm disk}+\Sigma_i \rho_{\rm cloud, i} + \rho_{\rm halo},
\ee 
where 
\be
\rho_{\rm cloud,i}=100\ e^{-((x-x_i)^2+(y-y_i)^2+z^2)/w_{\rm cloud,i}^2},
\ee
with 
$(x_i,y_i,w_{\rm cloud,i})=(5,0,1)$,  
$(-\sqrt{5},\sqrt{5},0.5)$, and $(-\sqrt{5},-\sqrt{5},0.5)$.  

Figure \ref{p-AtoB} shows a result of MHD wave convergence on such clouds. 
Because of the spherical convergence onto each cloud, the amplification of the wave flux is much stronger compared with convergence onto a ring. 
A fraction as high as $\sim 20$\% of the total released wave front from the nucleus is trapped by the largest cloud and focuses onto its center.
Figure \ref{p-AtoB-rot} shows a case when the disc and clouds are rotating with nearly constant velocity as indicated in figure \ref{Vrot}. 
The convergence of waves onto the clouds are essentially the same, but the focusing waves are deformed according to the differential rotation.

\subsection{Effect of vertical magnetic cylinder} 

Figure \ref{cylMag-AB} shows a result for the waves emitted from the nucleus (Sgr A) and propagate through a disc and vertical magnetic cylinder. 
The waves are reflected by the inner wall of the cylinder, and a fraction is trapped inside the wall and disc.
However, a remaining fraction penetrates the wall and propagate through the disc, and further converge onto the ring and molecular clouds. 

The magnetic cylinder, therefore, reflects the waves and confine some fraction inside the cylinder, and some fraction penetrates through the wall and converge onto the target clouds.
Although the magnetic cylinder somehow suppresses the efficiency of feedback from the wave source to the targets, the characteristic behavior of the waves are essentially the same as in the case without magnetic cylinder.

It may be noted that there appears an almost vacant region of waves around the cylinder's radius, where the waves propagate faster than the surrounding region because of the faster \Alf velocity.
This suggests that there is a quiet region of interstellar disturbances between Sgr A and molecular ring in the CMZ.

\section{Sgr B to A: Starburst to AGN}
\label{secsgrB}

\subsection{Ring to center}

In dense molecular clouds in the CMZ, star formation is triggered by the effective compression by the focusing MHD waves, which would be lead to active SF and may cause SB, if the compression is strong enough. 
Once active SF occurs, the preceding explosive phenomena such as supernova (SN) explosions, stellar winds, and expanding HII shells will produce various types of disturbances, which propagate through the CMZ and the galactic disk.

We assume a constant magnetic field with $B=1$ and gas density distribution as
\be 
\rho=\rho_{\rm disk}+\rho_{\rm nuc} +\rho_{\rm halo},
\ee
with 
$\rho_{\rm nuc}=100\ e^{-(r/r_{\rm nuc})^2},$
  and $r_{\rm nuc}=1$.

\begin{figure}
\begin{center}  
\includegraphics[width=7cm]{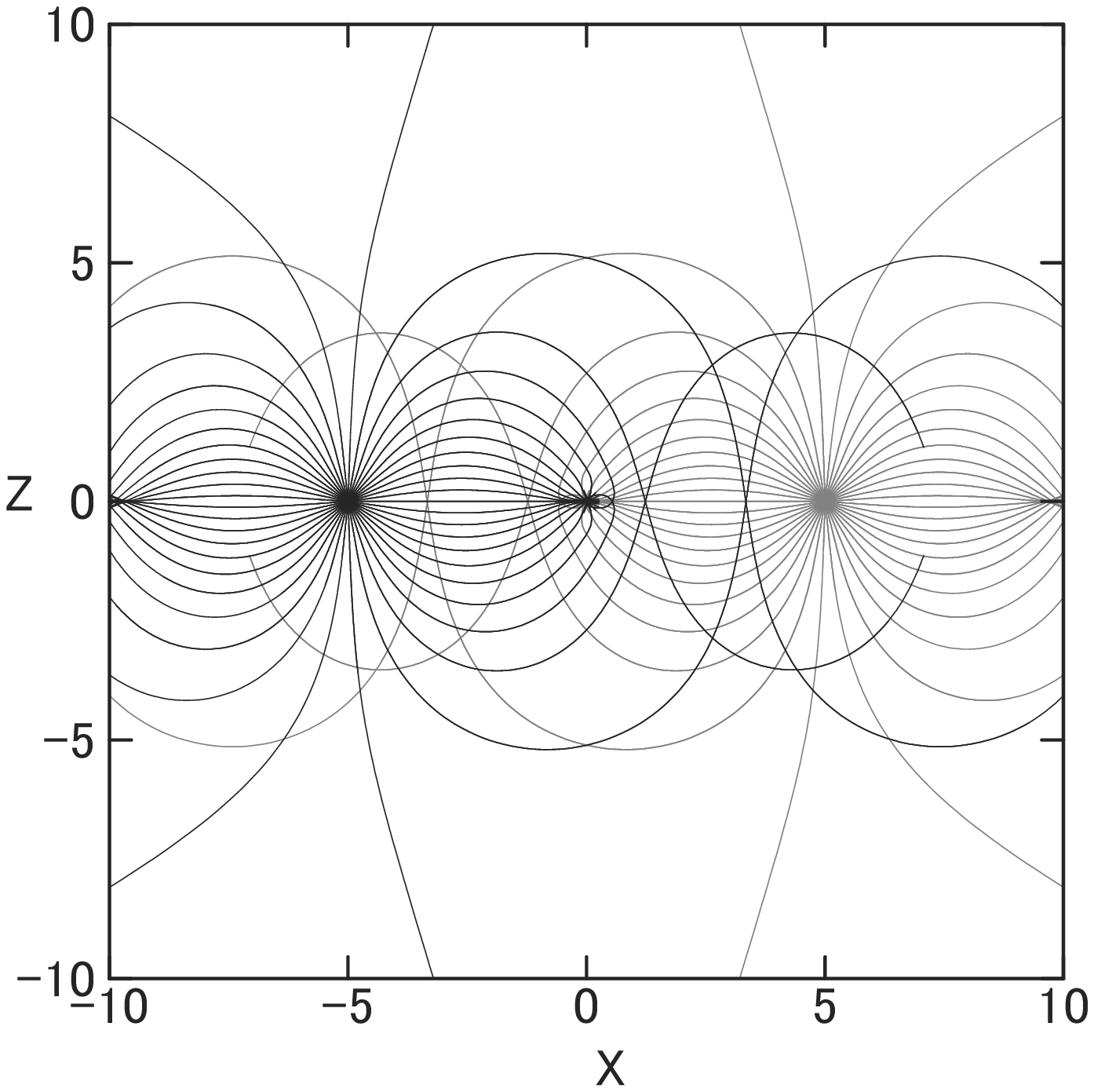}   
\end{center}
\caption{Paths of MHD waves excited in a ring of radius 5 focusing onto the nucleus.} 
\label{ring-center} 
	\end{figure}

Figure \ref{ring-center} shows propagation of MHD waves produced in a ring surrounding the GC, where the waves start from the ring at r=5 and propagate through the disc with plane parallel density distribution as sech z/1.0.
Black and red lines indicate ray paths starting from two points on the ring at $X=-5$ and 5, respectively.
The waves expands from the ring and are reflected by the sech disk. About a half of the rays focus on the Galactic Center at $X=0$.

The rays from every azimuth position on the whole ring focuses on the central one point at high efficiency.
After passing the center, the rays further propagate through the disc and focus again on the opposite side of the ring, and propagate further outward.

\subsection{Sgr B to A}

Although circum-nuclear SB is known to occur in a ring or donuts region surrounding the nucleus, the SF regions and molecular gas distribution are more or less clumpy, which is in fact clearly observed in the CMZ and associated SF sites 
\cite{Oka+1998,Oka+2012}. 
In the CMZ, the most active SF is observed in the Sgr B complex composed of HII regions and giant molecular clouds 
\cite{Hasegawa+1994}. Expanding shells of hot gas around Sgr B2 suggest strong wind or explosive events in the SF region
\cite{Martinpintado+1999}.

Radio continuum observations of Sgr B region has shown that the radio emission is a mixture of thermal and non-thermal emissions, indicating that the region contains high energy objects
\cite{Jones+2011} (and the literature therein). X-ray observations also suggest active events, heating the surrounding gas to high temperature
\cite{Koyama2018}. 
Thus, the SF activity in Sgr B will result in a variety of explosive and/or wind phenomena, including multiple supernova explosion. 	
We trace propagation of disturbances simulated by a spherical MHD wave originating in a remote site from the nucleus at $(x,y,z)=(5,0,0)$, mimicking an explosive event in the Sgr B SF complex. 
The gas density distribution is assumed to have the form as
\be
\rho=\sech \left(\frac{z}{h}\right)  
e^{ -\left( \frac{\varpi}{\varpi_{\rm disk}} \right)^2 }
+100 e^{- \left( \frac{r}{r_{\rm nuc}} \right)^2 } + 0.01,
\label{diskAB}
\ee
with $h=1$, $\varpi_{\rm disk}=10$ and $r_{\rm nuc}=1$, representing a sech disc of radius 10 and a nuclear high-density gas concentration of radius 1.0

\begin{figure*}
\begin{center}
{\bf [Sgr B $\Rightarrow$ Sgr A]}\\
\includegraphics[width=16cm]{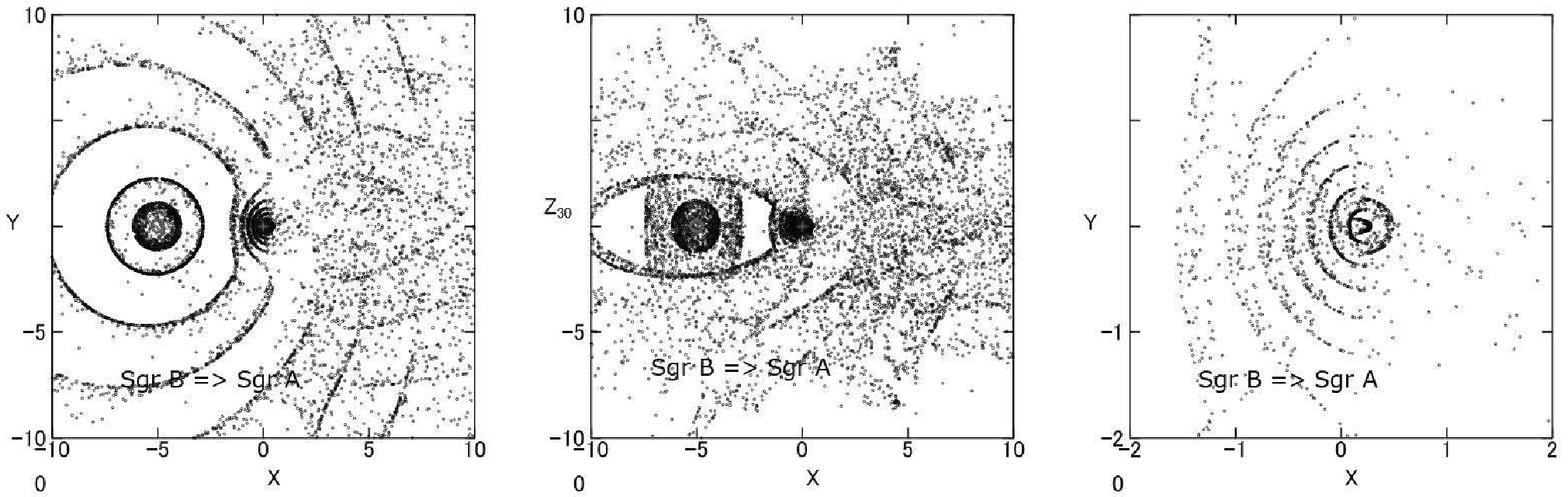}
\end{center}
\caption{(Left) Projection on the $(x,y)$ plane of SB-induced MHD wave by Sgr B (circle) focusing on the nucleus at Sgr A (cross). 
(Middle) Same, but seen from altitude at 30$\deg$. (c) Same as left panel, but enlarged near Sgr A.}
\label{sb_gc} 

\begin{center} 
{\bf [Sgr B in rotation $\Rightarrow$ Sgr A]}\\
\includegraphics[width=16cm]{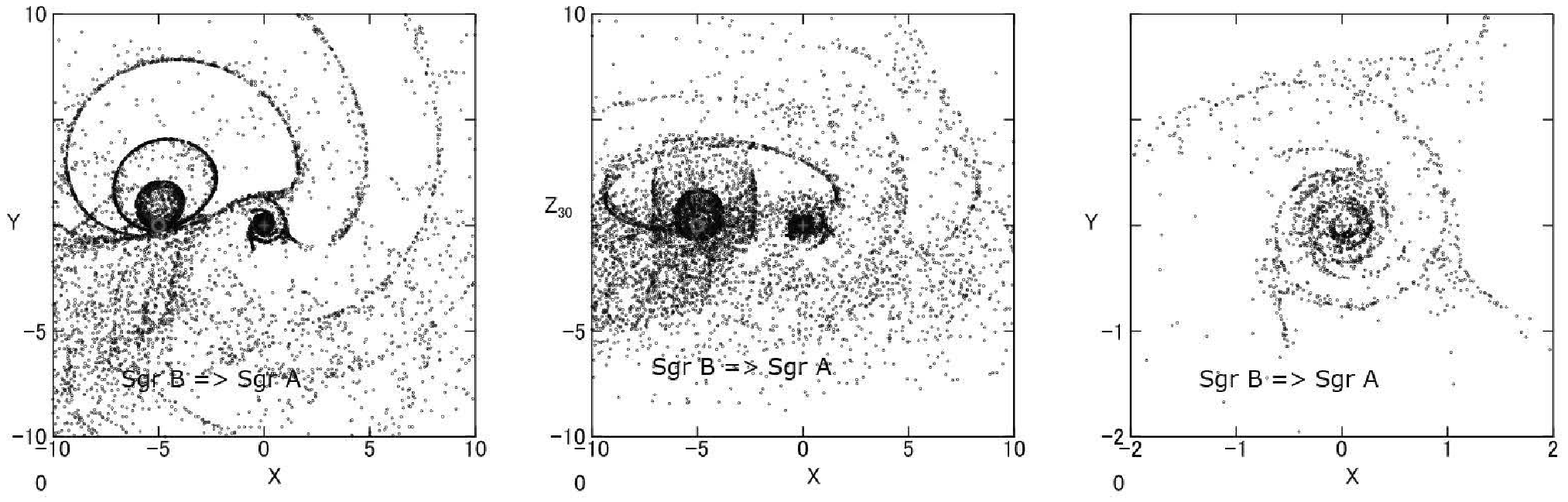}
\end{center}
\caption{Same as figure \ref{sb_gc}, but in a rotating disc with circular velocity as shown by figure \ref{Vrot}. }
\label{sb_gc_rot} 
	
\begin{center}    
{\bf [Sgr B, etc.  in rotation $\Rightarrow$ Sgr A]}\\
\includegraphics[width=16cm]{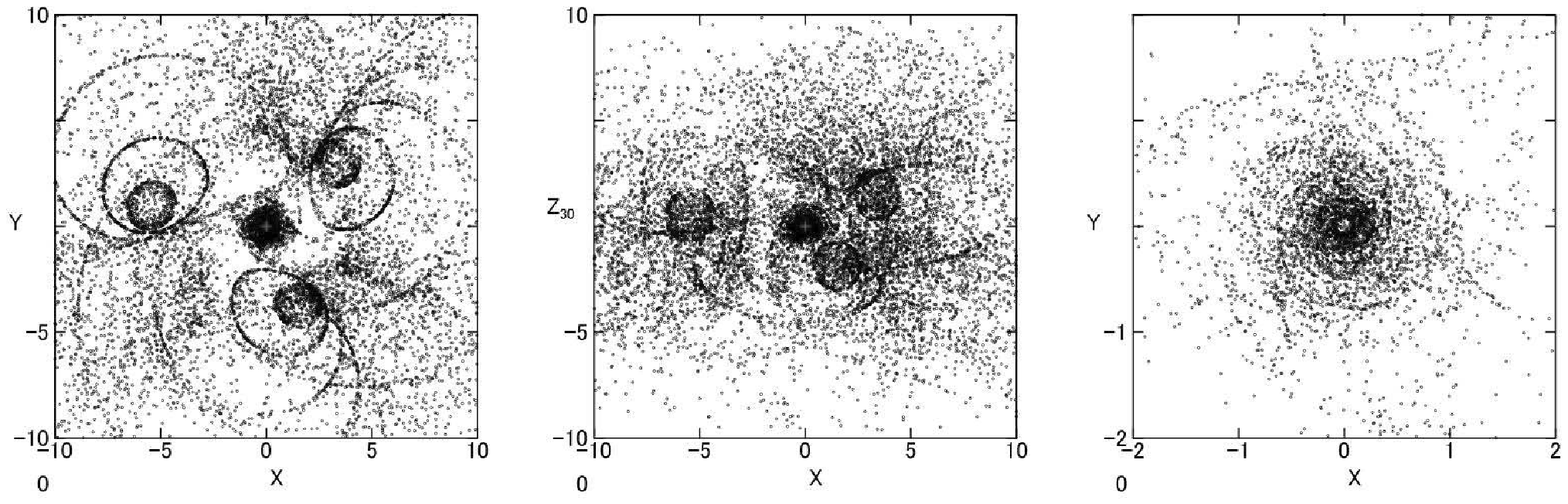}
\end{center}
\caption{Same as figure \ref{sb_gc_rot}, but the MHD waves are emitted  from three sources at $r=4, 4.5$ and 5.5, mimicking Sgr B, C etc.. }
\label{sb_multi_gc_rot}
\end{figure*}

Figures \ref{sb_gc}(a) and (b) show the wave front at t=1, 2, 4, 6, ... and 20 for non-rotating disk. The wave front expands spherically in the initial stage around the explosion place, mimicking disturbance from a SB site such as Sgr B. 
As the shell expands, it is deformed due to the sech disc to form cylindrical shape, and is further converges to the disk. 
A significant portion, about $sim 10$ percent, of the front facing the nucleus focuses onto the nucleus.

Figure \ref{sb_gc_rot}(c) and (d) are the same, but the disc is rotating at a constant velocity as shown in figure \ref{Vrot}. 
The wave source (Sgr B) moves along a circle at $r=5$ clockwise and the front expands in the rotating disk. 
Approximately the same portion of the front facing the nucleus as in case (a) focuses onto the nuclear gas cloud, while the front shape is deformed by the differential rotation. 
When the wave approaches the nucleus, the front shape attains almost spherical shape, focusing onto the center.

Figure \ref{sb_multi_gc_rot} shows the same in a rotating disk, but there are three wave sources at slightly different radii at $r=4$, 4.5 and 5, and different azimuth angles. This figure demonstrates the effective convergence of the waves from multiple SF (SB) sites onto the nucleus at Sgr A.

\begin{figure*}
\begin{center} 
{\bf [Sgr B etc. on a ring in rotation $\Rightarrow$ Sgr A]}\\
\includegraphics[width=16cm]{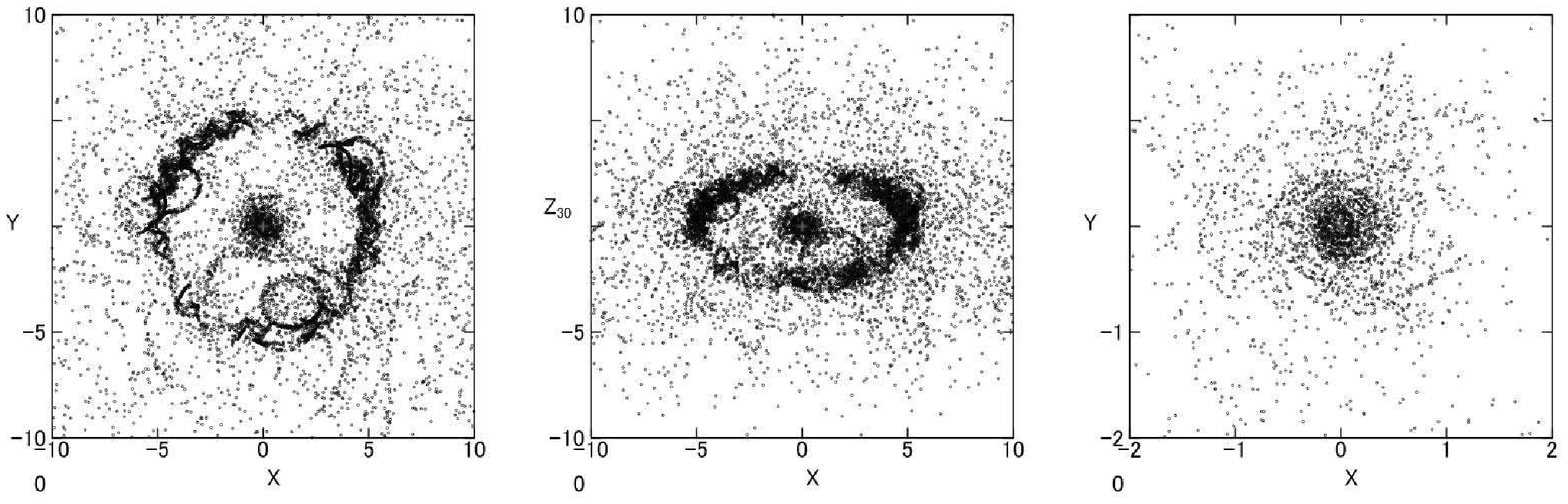}   
\end{center}
\caption{Same as figure \ref{sb_multi_gc_rot}, but a gas ring of radius 5 and half width 0.5 is added. Waves are either trapped in the ring, or  escape and focus on the nucleus (Sgr A). }
\label{sb_multi_ring_gc_rot} 
	\end{figure*}

Figure \ref{sb_multi_ring_gc_rot} shows the same, but the emitting sources (SF regions) are located on the edge of a molecular ring as observed as the 200 pc molecular ring \cite{Sofue1995} embedded in the disc given by equation (\ref{diskAB}). 
The ring's density is represented by 
\be
\rho_{\rm ring}=\rho_{\rm ring, 0}  e^{-((\varpi-5)^2+z^2)/w_{\rm ring}^2},
\ee 
where $\varpi_{\rm ring,0}=5$, half width $w_{\rm ring}=1$, and  $\rho_{\rm ring,0}=5$.
which is embedded in the disc given by equation (\ref{diskAB}).
A large fraction of the MHD waves are trapped in the ring, and a portion, about 10 percent, escapes from the ring and converges onto the nucleus.
The waves inside the ring efficiently converge to the center, propagating through the disc in differential rotation, and focus onto the nucleus, resulting in spherical implosion.  

It may be stressed that the major fraction of the MHD waves are trapped inside the ring, staying there almost without dissipation (equation (\ref{dissipation})).
The waves propagate along the ring, repeating oscillating focusing at wave length of $\sim 2w_{\rm ring}$ in the same way as in a magnetized filament \cite{Sofue2020a}. 

\subsection{Feedback through a magnetic cylinder}

\begin{figure*}
\begin{center} 
{\bf [Sgr B, etc. on a ring in rotation $\Rightarrow$ Sgr A through a magnetic cylinder]}\\
\includegraphics[width=16cm]{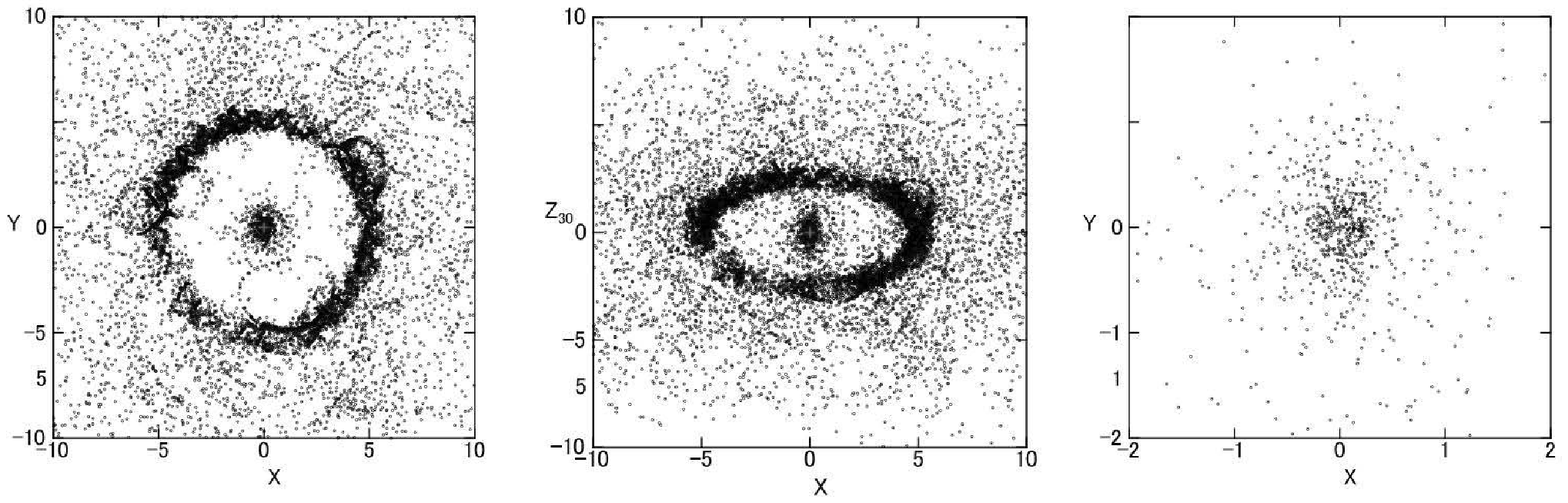}   
\end{center}
\caption{Same as figure \ref{sb_multi_ring_gc_rot}, but in the presence of a magnetic cylinder at $\varpi=3$. MHD waves from three SF regions (Sgr B etc.) near the molecular ring are reflected by the magnetic cylinder, while a fraction penetrates it and converges onto the nucleus (Sgr A). There appears a vacant area in the cylider, with the wave being rejected. Some self-feedback occurs back to the ring and clouds.}
\label{cylMag} 
	\end{figure*}

Figure \ref{cylMag} shows the result for MHD waves emitted from three SF regions near the molecular ring at $\varpi=5$ in the presence of a magnetic cylinder of radius 3.
A fraction of the waves are reflected by the magnetic wall, and propagate backward and trapped (absorbed) in the molecular ring.
Some other fraction penetrates through the magnetic cylinder, and converges onto the nucleus.
Thus, the magnetic cylinder somehow suppress the efficiency of feedback from Sgr B to Sgr A, although the essential behavior is about the same without magnetic cylinder.
Again, a vacant region appears coincident with the cylinder's radius due to the faster \Alf velocity inside.

\subsection{Efect of clumpiness}
\label{clumpiness}
	 
Although the CMZ is composed of such major structures as the central disc, molecular ring, giant clouds like Sgr B, and the nuclear core around Sgr A, it is also full of clumps of smaller scales and turbulence \cite{Morris+1996,Oka+1998}.
Such small structures cause fluctuations of the distribution of \Alf velocity, and disturb the smooth propagation of the MHD waves. 
While detailed discussion of magneto-hydrodynamic turbulence is beyond the scope of this paper, we here try to perform a simple exercise to examine how small scale fluctuations affect the MHD propagation by adding sinusoidal deformation against the background \Alf velocity distribution.
Namely, the \Alf velocity is multiplied by a factor of 
\be
f_{\rm clump}=1+q\ \sin (5x)\ \sin  (5y)\  \sin  (5z),
\ee 
with $q\sim 0.1$.
This equation represents $\Va$ variation with wavelength $\lambda \sim 0.6$ and peak-to-peak amplitude of 0.2 times the background, or peak-to-peak gas density of 0.4 times. 

\begin{figure*}
\begin{center} 
 {\bf [ Sgr B, etc.$\Rightarrow$ Sgr A; without/with backgound fluctuations ]}\\
\includegraphics[width=16cm]{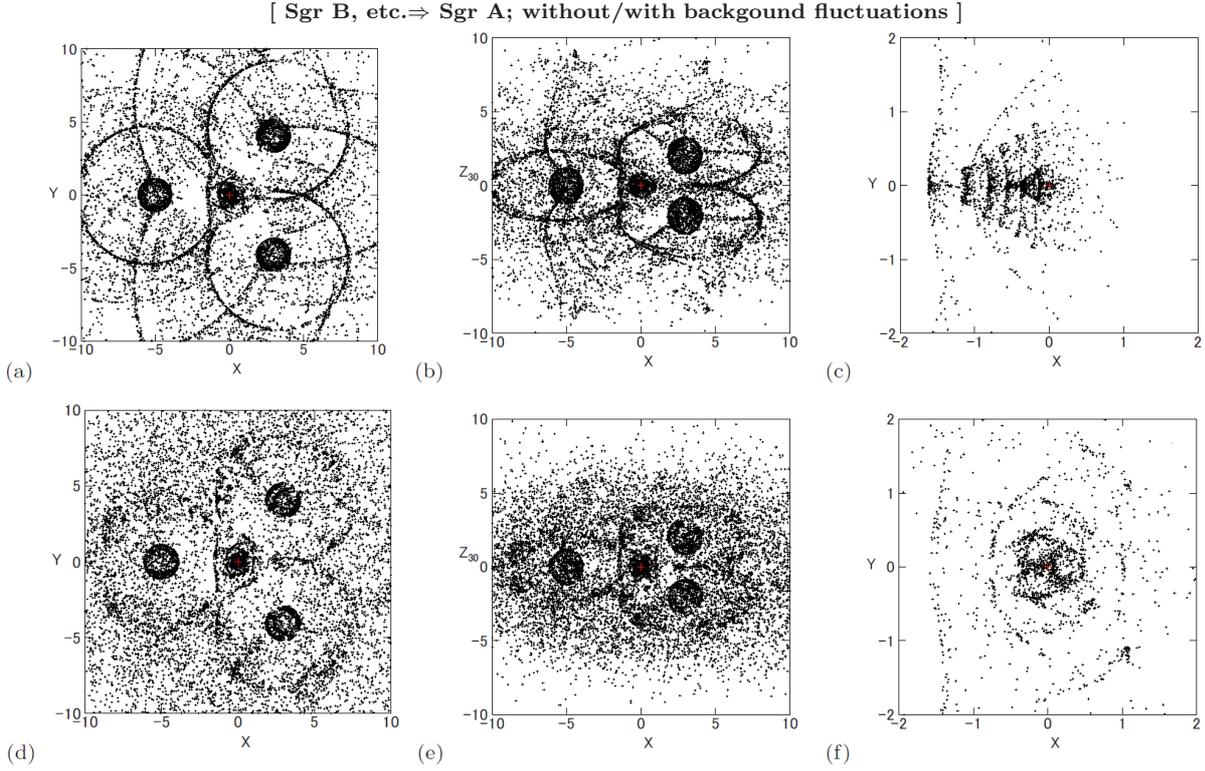}    
\end{center}
\caption{ Effect of clumpiness on the MHD waves from three sources (Sgr B, etc.) at $\varpi \sim 5$ without rotation, converging on Sgr A at the center. 
(a) \Alf velocity distribution is the same as for figure\ref{sb_multi_gc_rot}, and the waves from 3 sources are shown at  $t=1$, 4, 8, ..., 20 projected on the ($x,y$) plane.   
(b) Same, but projection seen from $30\deg$ above the galactic plane.
(c) Same as (a), but close up around Sgr A.
(d) to (f) Same as (a) to (c), respectively, but the background \Alf velocity is superposed by fluctuation multiplied by a factor of $f_{\rm clump}=1+0.1\times \ \sin(5x)\ \sin(5y)\ \sin(5z)$. } 
\label{clump}
\end{figure*}

 In order to abstract the effect, we examine a simple case where are put three major clouds (Sgr B etc.) without rotation and a nuclear core (Sgr A), and the waves are emitted at the surfaces of the three clouds.
In figure \ref{clump} we compare the results for no clumps (upper panels) and with clumps (lower panels).
Although the wave fronts are more diverted due to the scattering and diffraction of the ray paths by the clumps, the global behavior of the wave propagation does not much different between the two cases.
Focusing onto Sgr A also occurs at almost the same efficiency, while the directions of the focusing waves are more widely spread.
Therefore, we can conclude that the clumps disturb the front shape of the wave, but it has no significant effect on the global focusing on the nucleus and its efficiency.
The behavior is similar to a fluid flow collected by a deformed funnel into the central hole, regardless of the degree of deformation. 

\section{Discussion}
\label{secdiscussion}
\subsection{Echoing Feedback between Sgr A and B}

We have shown that MHD waves excited by the AGN in Sgr A converge onto the molecular clouds such as Sgr B in the CMZ. 
The waves focus onto the clouds' centers, and compress the gas to trigger star formation.
If the activity in Sgr A is strong and the wave amplitude is sufficiently large, the focusing wave will compress the cloud strongly, leading to starburst.

The thus triggered SF and SB produce expanding HII regions, SN explosions and stellar winds, which further excite MHD waves in the surrounding ISM. 
The major part of the MHD waves is trapped to the disc and ring, and some portion, $\sim 10-20$ percent, focus on the nucleus, or Sgr A, as spherically imploding compression wave. 

Thus, the activity of Sgr A (AGN) triggers the SF/SB in the CMZ (Sgr B), which excites another MHD waves that inversely converge back to Sgr A by the inward focusing. This boomerang-focusing cycle will continue until the circum-nuclear gas and CMZ are exhausted by star formation and out-flowing events such as winds and expanding shells.

\subsection{Solving the angular-momentum problem of AGN fueling}

The key problem about fueling AGN by accretion of cold gas is how to control the refusing forces by conservation of angular momentum and increasing magnetic pressure
\cite{Umemura+1997,Shlosman+1989,Krolik+1990,Thompson+2005,Jogee2006}.
Such problem can be solved by the present model, because the MHD wave propagates as local and temporary enhancement of the magnetic pressure associated with gas compression, surfing the rotating disc without transporting angular momentum.
Namely, the convergence onto the nucleus occurs without global change in the angular momentum and magnetic field.  
Thus, the MHD wave focusing can produce spherical implosion onto the nucleus without suffering from refusing forces by conservation of angular-momentum as well as magnetic pressure.  

\subsection{Energetics}

The kinetic energy released by the starburst in the CMZ can be approximately estimated by the supernova (SN) rate and SF rate in the GC. 
High excess of small-diameter SNRs in the GC \cite{Gray1994} suggests that the SN rate per volume is higher than that in the galactic disc. 
 The observed SF rate of $\sim 0.1 \Msun$ y$^{-1}$ in the CMZ \cite{Barnes+2017,YZ+2009} indicates massive-star birth rate of $\sim 10^{-3} \times 10 \Msun$ (OB stars) y$^{-1}$ $\sim 10^{-2}\Msun$ y$^{-1}$, \red{corresponding to $\sim 10^{-3}$ SNe y$^{-1}$.}  
Then, the injection rate of the kinetic energy into the CMZ by SNe is on the order of $L_{\rm SN}\sim \eta E_{\rm SN} dN_{\rm SN}/dt \sim 10^{39}$ ergs s$^{-1}$, where $E_{\rm SN}\sim 10^{51}$ ergs and $\eta\sim 0.03$ is the fraction of kinetic energy. The injected kinetic energy by SNRs finally fades away and merge into the ISM of CMZ, and excites small amplitude MHD waves. 
As the simulation showed, a significant fraction, $\sim 0.1$, of thus created waves converges onto the nucleus by the focusing effect. 
\red{This results in an implosive injection of kinetic energy in the form of compression MHD waves at a rate of $L_{\rm kin} \sim 10^{38}$ ergs s$^{-1}$ into a small focal area around Sgr A$^*$ (figure \ref{sb_gc}). }

 Since the problem of angular momentum has already been solved as in the previous subsection, this kinetic energy is directly spent to promote the accretion of the circum-nuclear gas towards the center, overcoming the gravitational barrier. 
The accretion rate is related to the injecting kinetic energy luminosity as
\be 
L_{\rm kin}\sim \dot{M} \( \frac{GM_\bullet}{r} \),
\ee 
or 
\be 
\dot{M}
\sim 3.6 \(\frac{L_{\rm kin}}{10^{40}{\rm erg}} \) \( \frac{r}{1 {\rm pc}}\) \(\frac{M_\bullet}{10^6 \Msun}\)^{-1} \ \Msun \ {\rm y}^{-1},
\ee 
where $M_\bullet$ is the mass of the central massive object and $r$ is the radius at which the accretion is proceeding.
\red{For above luminosity, this reduces to $\dot{M}\sim 0.01(r/1{\rm pc})$, if we assume that the convergence is so efficient that the focusing occurs into a small volume around the central massive black hole of mass $M_\bullet \sim 4\times 10^6 \Msun$ at Sgr A$^*$ \cite{Genzel+2010}.
This yields $\dot{M}\sim 10^{-6}\Msun$ y$^{-1}$ for $r\sim 10^5 R_{\rm S}$ with  $R_{\rm S}$ being the Schwaltzschild radius, and would be compared with the rate of a few $10^{-6}\Msun$ y$^{-1}$ estimated for Sgr A$^*$  \cite{Cuadra+2005,Yuan+2014},
although it is beyond the scope of the model how the accretion can further reach this radius from the presently simulated scale of several pc.
} 

\subsection{High efficiency compression by spherical implosion}

The present MHD calculation by solving the Eikonal equations cannot treat non-linear growth of waves and absolute amplitude.
However, we may speculate that the wave will grow rapidly as it focuses on the focal point, where the amplitude increases inversely proportional to the spherical surface of the wave front.
Such implosive focusing will further cause strong and efficient feeding of compressed gas onto the focal point such as the nucleus or dense cloud's center.

An advantage of the present model is its minimal energy requirement. 
The released energy from the AGN in the form of MHD waves propagates the GC disc without dissipation as estimated by equation (\ref{dissipation}).
Almost all fraction of the waves are trapped inside the GC disc and CMZ, and converges onto the clouds, where the waves focus on focal points.
Even weak disturbances in the form of MHD waves are collected by the clouds and largely amplified, causing spherical implosion toward the focal points. 

\subsection{Shock waves in explosion phase vs MHD waves in quiet phase}
	 
There have been various models to explain the radio, X-ray and $\gamma$-ray bubbles and shells from the Galactic Center by energetic explosions associated with strong shock waves
\cite{Crocker2012,Kataoka+2018}.
Such explosive phenomena will significantly influence and change the structure of the galactic disk, and may work to suppress star formation, rather than to trigger, by blowing off the gas from the disk.

Such violent phenomena make contrast to the model presented here of triggering the activities and star formation by focusing of MHD waves.
Opposite two different situations may be possible to occur, if there are two different activity phases, strong and weak, in the nucleus as follows.

One phase is composed of energetic explosions associated with giant supersonic shells and jets expanding into the halo, which may blow off the surrounding interstellar medium into the halo.
The other is weak and silent phase between the strong ones, during which weak disturbances such as MHD waves are emitted gently and constantly, and trigger the SF by focusing implosions. 
Although weak, the latter will last much longer than the strong phase, so that the nucleus (Sgr A) may be regarded as a constant supplier of the triggering waves.
 
 \subsection{Larger-scale feedback in the entire Galaxy}  
 The present feedback mechanism can be extended to larger scale feedback of explosive energy to the disc and nucleus in the entire Milky Way. 

 MHD waves produced at the nucleus converge not only onto the CMZ, but also penetrate it and propagate through the entire disc of the Milky Way because of the small dissipation rate.
This may cause further convergence onto the spiral arms, molecular and HI clouds, and would act to trigger implosive compression of the clouds, leading to star formation. 
Stronger shock waves from the nucleus expanded into the halo make giant shells and bubbles. 
When the shells fade out in the halo, sound and MHD waves are excited in the halo.
Such waves are then reflected by the upper halo, and converge onto the galactic disc and further to interstellar clouds.
Thus, most of the released kinetic energy at the nucleus (Sgr A) is trapped inside the Milky Way, and fed back to interstellar clouds, triggering there subsequent star formation. 

 Similarly, MHD waves excited in the galactic disc by SF and SNe propagate in the disk, and a significant portion globally converges to the galactic center, triggering AGN.
Again, the efficiency of wave trapping and focusing is so high that a significant fraction of the kinetic energy released by SF is fed back to the GC.
The convergence of these waves to the nucleus will continue as long as SF activity continues in the Galaxy. 

\vskip 5mm
\section{Summary}

We have traced the propagation of fast-mode magneto-hydrodynamic (MHD) compression waves in the Galactic Center (GC). 
It was shown that the waves produced by the activity in the nucleus (Sgr A) focus on the molecular ring and clouds in the CMZ, which will trigger starburst. 
As feedback, MHD disturbances induced by SF activity or starburst propagate backward to the nucleus, and focus on the cloud around Sgr A.
This further enhance implosive compression to cause nuclear activity.
The present model, thus, solves the most important problem of the angular momentum in the AGN fueling mechanism.  
The AGN (Sgr A) and starburst (Sgr B) trigger each other through echoing focusing of MHD waves, which realises mutual trigger at high efficiency and minimal energy requirement.
The present idea and method would also be applied generally to insight into the detailed mechanism of the AGN-SB connection in external galaxies.
 
\vskip 5mm
\noindent{\bf Data availability} There are no data available on line.

\noindent {\bf Acknowledgements}: The calculations were performed at the Astronomical Data Center (ADC) of the National Astronomical Observatory of Japan (NAOJ). The author would like to thank the anonymous referee for the valuable comments.


\begin{appendix}

\section{Eikonal equations}
\def\pr{p_r} \def\pth{p_\theta} \def\pphi{p_\phi} \def\cot{ {\rm cot} }  \def\d{\partial} \def\f{\frac}

 The Eikonal equations describing the propagation of a MHD wave packet are given as follows \cite{Uchida1970,Sofue1978}.
\be \f{dr}{dt}=V \f{\pr}{p}, \ee
\be \f{d\theta}{dt}=V \f{\pth}{rp}, \ee 
\be \f{d\phi}{dt}=V \f{\pphi}{rp\  \sin \theta}+\Omega, \ee 
\be \f{d\pr}{dt}=-p\f{\d V}{\d r} +\f{V}{rp}(\pth^2+\phi^2), \ee
\be \f{d\pth}{dt}=-\f{p}{r} \f{\d V}{\d \theta} 
- \f{V}{rp}(\pth \pr-\pphi^2 \ \cot \ \theta), \ee 
\be \f{d\pphi}{dt}=-\f{p}{\sin \theta} \f{\d V}{\d \phi} - \f{V}{rp}(\pphi \pr + \pphi \pth \ \cot \ \theta), \ee 
where
$V=V(r,\theta,\phi)=V(x,y,z)$ is the \Alf velocity, the vector $\bm{p}=(\pr, \pth, \pphi)={\rm grad}\  \Phi$ is defined by the gradient of the eikonal $\Phi$, $p=(\pr^2+\pphi^2+\pth^2)^{1/2}$, $(r,\theta, \phi)$ and $(x,y,z)$ are the polar and Cartesian coordinates, and $\Omega$ is the angular velocity of the ambient material around the Galactic rotation axis ($z$ axis). 

\end{appendix} 

\end{document}